\documentclass[3p]{elsarticle}

\usepackage{hyperref}
\usepackage{amsmath}
\usepackage{amssymb}
\usepackage{amsthm}
\usepackage{mathrsfs}
\usepackage{cleveref}
\usepackage{xspace}
\usepackage{calc}  
\usepackage[shortlabels]{enumitem}

\journal{Information Processing Letters}

\newcommand{\WRP}{\textsc{Waypoint Routing Problem}\xspace} 
\newcommand{\W}{W} 
\renewcommand{\c}{\kappa} 
\newcommand{\w}{\omega} 

\newcommand{\mincup}{%
  \mathbin{\ooalign{$\cup$\cr\hss\raisebox{0.5ex}{\scriptsize $\downarrow$}\hss}}}%
\newcommand{\bigmincup}{%
\mathop{\ooalign{$\displaystyle\bigcup$\cr\hidewidth{\Large$\downarrow$}\hidewidth\cr}}}%
\newcommand{\N}{\mathbb{N}} 
\newcommand{\WPS}{\Pi(U)\times\N}
\newcommand{\NP}{$\mathbf{NP}$\xspace} 
\newcommand{\FPT}{$\mathbf{FPT}$\xspace} 

\newcommand{\T}{\mathcal{T}} 
\DeclareMathOperator{\tw}{tw} 
\newcommand{\A}{\mathcal{A}} 
\newcommand{\Pttn}{\mathcal{P}} 
\newcommand{\Qttn}{\mathcal{Q}} 

\newcommand{\opt}{\mathtt{opt}}
\newcommand{\rmc}{\mathtt{rmc}}
\newcommand{\ins}{\mathtt{ins}}
\newcommand{\shift}{\mathtt{shift}}
\newcommand{\glue}{\mathtt{glue}}
\newcommand{\proj}{\mathtt{proj}}
\newcommand{\join}{\mathtt{join}}
\newcommand{\reduce}{\mathtt{reduce}}

\theoremstyle{definition}
\newtheorem{definition}{Definition}
\newtheorem{notation}{Notation}
\theoremstyle{plain}

\newtheorem{lemma}{Lemma}

\newtheorem{theorem}{Theorem}

\usepackage[mathlines]{lineno}
\modulolinenumbers[1]
\usepackage{etoolbox} 

\newcommand*\linenomathpatchAMS[1]{%
   \expandafter\pretocmd\csname #1\endcsname {\linenomathAMS}{}{}%
   \expandafter\pretocmd\csname #1*\endcsname{\linenomathAMS}{}{}%
   \expandafter\apptocmd\csname end#1\endcsname {\endlinenomath}{}{}%
   \expandafter\apptocmd\csname end#1*\endcsname{\endlinenomath}{}{}%
}

\expandafter\ifx\linenomath\linenomathWithnumbers
   \let\linenomathAMS\linenomathWithnumbers
\patchcmd\linenomathAMS{\advance\postdisplaypenalty\linenopenalty}{}{}{}
\else
   \let\linenomathAMS\linenomathNonumbers
\fi

\linenomathpatchAMS{gather}
\linenomathpatchAMS{multline}
\linenomathpatchAMS{align}
\linenomathpatchAMS{align*}
\linenomathpatchAMS{equation}
\linenomathpatchAMS{equation*}
\linenomathpatchAMS{alignat}
\linenomathpatchAMS{flalign}

\begin{document}

\begin{frontmatter}

	\title{Waypoint Routing on Bounded Treewidth Graphs\fnref{cc}}

	\author{\v{S}imon Schierreich\fnref{fn1,fn2}}
	\author{Ond\v{r}ej Such\'y\fnref{fn1}}
	\address{Department of Theoretical Computer Science, Faculty of Information Technology,\\ Czech Technical University in Prague, Prague, Czech Republic}
	\fntext[fn1]{The author acknowledges the support of the OP VVV MEYS funded project CZ.02.1.01/0.0/0.0/16\_019/0000765 ``Research Center for Informatics''.}
	\fntext[fn2]{This work was supported by the Grant Agency of the Czech Technical University in Prague, grant No. SGS20/208/OHK3/3T/18, and by the
	Student Summer Research Program 2019 of FIT CTU in Prague.}
	\fntext[cc]{© 2021. This manuscript version is made available under the CC-BY-NC-ND 4.0 license \url{https://creativecommons.org/licenses/by-nc-nd/4.0/}}

	\begin{abstract}
		In the \WRP one is given an undirected capacitated and weighted graph $G$, a source-destination pair $s,t\in V(G)$ and a set $\W\subseteq V(G)$, of \emph{waypoints}. The task is to find a walk which starts at the source vertex~$s$, visits, in any order, all waypoints, ends at the destination vertex $t$, respects edge capacities, that is, traverses each edge at most as many times as is its capacity, and minimizes the cost computed as the sum of costs of traversed edges with multiplicities. 
		We study the problem for graphs of bounded treewidth and present a new algorithm for the problem working in $2^{\mathcal{O}(\tw)}\cdot n$ time, significantly improving upon the previously known algorithms. 
		We also show that this running time is optimal for the problem under Exponential Time Hypothesis. 
	\end{abstract}


\end{frontmatter}


\section{Introduction}

We study the \WRP (WRP), which can be formally defined as follows:
The input of the problem is an undirected, simple, capacitated and weighted graph $G = (V,E,\c,\w)$, where $\c\colon E \to \N$ and $\w\colon E \to \N$ represent edge capacities and edge weights, respectively, a source-destination pair $s,t\in V$, and a set $\W\subseteq V$, of \emph{waypoints}. 
A solution to the instance is a walk $R$ which starts at the source vertex $s$, visits, in any order, all waypoints $w\in\W$, ends at the destination vertex $t$, and respects edge capacities, that is, for each edge $e\in E$, the walk $R$ traverses $e$ at most $\c(e)$ times. 
Note that the order of the waypoints along the walk is not prescribed by the input and may be selected arbitrarily to suit the solution. 
Among all walks satisfying the condition above, we aim to minimize the cost of $R$, computed as a sum of weights of all edges in solution with multiplicities.

The problem is motivated by modern networking systems connecting distributed network functions, often composed of middleboxes, possibly virtualized, such as service chaining~\cite{SouleBMPKSF18}, hybrid software defined networks~\cite{CaniniFLSS14}, or segment routing~\cite{FilsfilsNPCF15}.

Study of (this variant of) the \WRP was initiated by Amiri et al.~\cite{AmiriFS18}. 
Inspired by the result of Rost et al.~\cite{RostDS19}, who showed that many natural network topologies such as backbone transit and wide area networks have small treewidth (i.e., they are in certain sense similar to trees), Amiri et al.~\cite{AmiriFS18} analyzed parameterized complexity of the problem with respect to the treewidth (denoted $\tw$) of the input graph. 
They presented a dynamic programming algorithm with running time $n^{\mathcal{O}(\tw^2)}$. %

As the running time of this algorithm seems rather impractical, it raises a question, whether it can be improved to say $n^{\mathcal{O}(\tw)}$ or even $f(\tw)\cdot n^{\mathcal{O}(1)}$ for a suitable computable function $f$. %

\paragraph{Our contribution} In this paper  we answer this question in the affirmative. In particular, we show that the \WRP can be solved in $2^{\mathcal{O}(\tw)}\cdot n$ time, that is, single exponential linear time, significantly improving upon the running time of the algorithm of Amiri et al.~\cite{AmiriFS18}. We also show that the running time of our algorithm is asymptotically optimal, assuming the popular Exponential Time Hypothesis of Impagliazzo and Paturi~\cite{ImpagliazzoP01}.

\paragraph{Related work} Apart from the above mentioned result, Amiri et al.~\cite{AmiriFS18} also showed that WRP admit a randomized $2^k\cdot n^{\mathcal{O}(1)}$-time algorithm, where $k$ is the number of waypoints, which can be turned into a deterministic polynomial time algorithm when $k  
= \mathcal{O}( (\log\log n)^{1/10})$. They also showed that the problem is \NP-hard on any class of graphs of maximum degree 3 on which \textsc{Hamiltonian Cycle} is \NP-hard. By a simple padding argument they showed that this also holds when $k=\mathcal{O}(n^{1/r})$ for any constant $r \ge 1$.
We are not aware of any result concerning approximability of the problem.

More research was concerned with the \textsc{Ordered WRP}, where the exact order in which the waypoints must be visited is given as a part of the input~\cite{AmiriFJS18,AmiriFJP18}. Amiri et al.~\cite{AmiriFJP18} showed that this problem can be solved in polynomial time on outerplanar graphs, while it is \NP-hard on graphs of treewidth $3$. Furthermore, it is \FPT with respect to $k$ and $tw$ combined. They also studied \textsc{Ordered WRP} on special classes of directed graphs. 

Amiri et al.~\cite{AmiriFJS18} extended these results by showing that the \textsc{Ordered WRP} is \FPT with respect to $k$ alone and that even the question whether any walk exists, let alone its weight, is \NP-hard for an unbounded number of waypoints. They further studied the problem for the case of one waypoint and showed that it is \NP-hard on directed graphs, or if the demanded flow changes in the waypoint, or if we pose some other constraints on the solution walk. 

The WRP is also somewhat related to the \textsc{Shortest disjoint paths} problem, where the task is to find $k$ disjoint paths between a specified pair of vertices such that the sum of lengths of the paths is minimized. Interestingly, for this problem, even for $k=2$, only a randomized polynomial time algorithm is known, but no deterministic~\cite{BjorklundH19}. Another well studied related problem is $k$-\textsc{Cycle}, where the task is to find a cycle through $k$ specified vertices~\cite{BjorklundHT12}. The problem has also similarities with \textsc{Subset TSP}, where the task is to find  the shortest tour through a specified subset of vertices~\cite{KleinM14}.

A somewhat opposite approach of placing the middleboxes or service chains in a software defined network in such a way that both load and routing constraints are fulfilled is also studied~\cite{MaBPP19,KutielR19}.

\section{Preliminaries}

We follow the basic notation of graph theory~\cite{diestel2017} and parameterized complexity theory~\cite{CyganFKLMPPS15}.

For a set $U$, we say that a family of its subsets $\Pttn \subseteq 2^U \setminus \emptyset$ is \emph{a partition} of $U$ if we have $\bigcup_{A \in \Pttn}A=U$ and for every two distinct $A,B \in \Pttn$ we have $A \cap B=\emptyset$.

\paragraph{Graph theory} An undirected \emph{multigraph} is a pair $G=(V,E)$, where $V$ is a finite set of vertices and $E$ is a set of edges, together with map $\eta\colon E\to[V]^2$ assigning to every edge two vertices --- its \emph{endpoints}. If there are no $e_1,e_2\in E$ such that $e_1 \neq e_2$ and $\eta(e_1)=\eta(e_2)$, i.e., parallel edges, we say that $G$ is a \emph{simple graph}. To simplify the notation we abbreviate the fact that $u$, $v$ are endpoints of $e$ to $e=\{u,v\}$, despite that the edge is no longer determined uniquely.

A \emph{walk} $R$ in (multi)graph $G$ is nonempty alternating sequence $v_1,e_1,v_2,\ldots,e_{k-1},v_k$ such that $\forall i \in [k]\colon v_i\in V$ and $e_i = \{v_i,v_{i+1}\}$. If it holds that $\forall i,j \in [k]\colon v_i=v_j \iff i=j$, then $R$ is a \emph{path}.

\paragraph{Parameterized complexity} Let $\Sigma$ be a fixed, finite alphabet. A language $L\subseteq \Sigma^*\times\N$ is called \emph{parameterized problem} and $(x,k)\subseteq\Sigma^*\times\N$, where $x$ is called \emph{input} and $k$ is a \emph{parameter}, is an \emph{instance} of problem $L$. Moreover an instance $(x,k)$ is called $yes$-instance if and only if $(x,k)\in L$.

A parameterized problem $L$ is called \emph{fixed-parameter tractable} (FPT) if there exists algorithm $\mathbb{A}$ correctly deciding whether $(x,k)\in L$ in time bounded by $f(k)\cdot|(x,k)|^c$, where $f\colon\N\to\N$ is a computable function and $c\in\N$ is a constant. The complexity class containing all such problems is called \FPT.

\paragraph{Tree decomposition and treewidth} A \emph{tree decomposition} of a graph $G=(V,E)$ is a pair $\T = \left(T, \beta\right)$ where $T$ is a tree and $\beta\colon V(T)\to2^V$ is a function that associates to each node $t\in V(T)$ a vertex subset $B_t\subseteq V(G)$, called a \emph{bag}, such that
\begin{enumerate}
 \item for each vertex $v\in V$, there is a node $t\in V(T)$ such that $v\in B_t$,
 \item for each edge $\left\{u,v\right\}\in E$, there is a node $t$ of $T$ such that $\left\{u,v\right\}\in B_t$,
 \item for each $v\in V$ the nodes $t$ such that $v\in B_t$ induce a connected subtree of $T$.
\end{enumerate}
The \emph{width} $w(\T)$ of a tree decomposition $\T$ is $\max_{t\in V(T)}\left\{|B_t| - 1\right\}$. The \emph{treewidth} of a graph $G$, denoted $\tw(G)$, is a minimum width of a decomposition over all tree decompositions of $G$.

Bodlaender et al.~\cite{BodlaenderDDFLP16} proved that a tree decomposition of a graph $G$ of width $\mathcal{O}(\tw)$ can be computed in $2^{\mathcal{O}(\tw)}\cdot n$ time.

\section{Graph unification}

In this section we present some easy modifications on the input instance in order to simplify it. Many of them are rather folklore for similar problems and were also used by Amiri et al.~\cite{AmiriFS18} or Klein and Marx~\cite{KleinM14}.

As a walk can never leave the connected component containing $s$, we can delete all other connected components of the input graph. If this also deletes $t$ or any waypoint, we can immediately answer no. As this can be performed in linear time, we can assume that the input graph is connected. As adding $s$ and $t$ into $\W$ does not change the problem, we assume $s,t \in \W$.

If $s\neq t$, then we introduce a new vertex $v$ to $G$ and connect it by edges to both $s$ and $t$. We set both the capacity and the cost of these two edges to $1$. It is easy to observe that there is a solution walk of cost $w$ in the original network if and only if there is solution walk from $v$ to $v$ through all waypoints of cost $w+2$ in the modified network. This also increases the treewidth of the graph by at most $1$, since we can add $v$ to every bag of the decomposition for $G$. Hence, we assume that $s=t$. 

Next, we interpret the task in terms of searching for appropriate Eulerian subgraphs of a suitable multigraph. 
Such an approach was first proposed by Cosmadakis and Papadimitriou~\cite{CosmadakisP84}.
Let us first define this multigraph.

Let $I=(G,\c,\w,s,t,\W)$ be an instance of WRP such that $G$ is connected and $s=t \in \W$, let $G'$ be a multigraph obtained by replacing each edge $e\in E(G)$ by $\c(e)$ parallel edges $e^1,\ldots,e^{\c(e)}$, $\w'(e^i)=\w(e)$ for every $e\in E(G)$ and $i \in \{1, \ldots, \c(e)\}$, and $\c'$ be equal to $1$ for each edge $e\in E(G')$.

\begin{lemma}\label{lem:solution_subgraphs}
	There is an $s$-$t$ walk of total cost $w$ in $G$ passing through all waypoints in $\W$ and obeying the capacities if and only if there is a \emph{connected} subgraph (submultigraph) $H$ of $G'$ of cost $w$ such that $\W \subseteq V(H)$ and $H$ is even (every vertex of $H$ has an even degree in $H$). 
\end{lemma}
\begin{proof}
	Given a walk $R$ in $G$ we can transform it into a walk $R'$ in $G'$ by replacing, for each edge $e$ occurring in $R$, its first occurrence by $e^1$, the second occurrence by $e^2$, etc. As $R$ obeys the capacities, this is possible and the cost of $R'$ is the same as the cost of $R$. Note that $R'$ traverses each edge at most once, i.e., it is a trail. Now let $V(H)$ and $E(H)$ be the set of vertices and edges traversed by $R'$, respectively. Then $R$ is an Eulerian trail for $H$, i.e., $H$ is connected and even. As $R'$ traverses all vertices of $W$, we have $\W \subseteq V(H)$ and the cost of $H$ is exactly the cost of $R'$ which is the cost of $R$.

	For the other direction, if we have the subgraph $H$, it is enough to take an Eulerian trail for this subgraph starting and ending in $s=t \in W$ to obtain the desired walk.
\end{proof}

Observe that if $H$ is the desired subgraph of $G'$ and it uses some edge more than twice (more than two parallel edges), then if we reduce the multiplicity of the edge by two, we obtain another subgraph with the desired properties. As each edge has a positive cost, this new subgraph will have a lower total cost. It follows, that in an optimal solution, each edge is traversed at most twice. Hence we introduce into $G'$ only at most two copies of each edge, without affecting the set of optimal solutions.

We call an instance \emph{unified}, if the (multi)graph is connected, has unary capacities, parallel edges have the same weight, there are at most two parallel edges between any pair of vertices, $s=t$, and $s \in \W$. Henceforward we will assume that the input instance of WRP is unified.

Amiri et al~\cite{AmiriFS18} explicitly asked whether WRP can be expressed in Monadic Second Order Logic (MSOL). It is straightforward to express the task of searching for the least cost Eulerian subgraph of a unified instance containing all the waypoints in a suitable extension of MSOL that allows counting the sizes of sets modulo 2 and finding the subset of interest of least cost in an element-weighted universe. As the classic result of~Courcelle \cite{Courcelle90} holds even for such extensions~\cite{CourcelleM93}, this shows that WRP is \FPT when parameterized by treewidth. Nevertheless, the running time of the algorithm implied by this meta-theorem is far from optimal. Therefore, we present a specialized dynamic programming algorithm in the next section.

\section{Dynamic programming algorithm}

To simplify the description of the algorithm, we will describe the computation in terms of a \emph{nice tree decomposition} of the underlying graph $G$. 

\begin{definition}[{Nice tree decomposition,~\cite[pp. 161 and 168]{CyganFKLMPPS15}}]\label{def:nice_tree_decomposition}
    A tree decomposition $\T=(T,\beta)$ is called \emph{nice} if and only if
    \begin{enumerate}
     \item $T$ is rooted in one vertex $r$ and $B_r = \emptyset$,
     \item $B_l = \emptyset$ for every leaf $l$ of $T$,
     \item Every internal node $t\in V(T)$ is of one of the following types.
     \begin{itemize}
      \item \emph{Introduce vertex node}: an internal node $t$ with exactly one child $t'$ such that $B_t = B_{t'} \cup \left\{v\right\}$ for some vertex $v \not\in B_{t'}$,
      \item \emph{Introduce edge node}: an internal node $t$, labeled with an edge $\left\{u,v\right\}\in E(G)$ such that $u,v \in B_t$, and with exactly one child $t'$ such that $B_t = B_{t'}$,
      \item \emph{Forget node}: an internal node $t$ with exactly one child $t'$ such that $B_t = B_{t'}\setminus\left\{v\right\}$ for some $v\in B_{t'}$,
      \item \emph{Join node}: a node $t$ with two children $t_1,t_2$ such that $B_t = B_{t_1} = B_{t_2}$.
     \end{itemize}
    \item For each edge $e$ there is exactly one node labeled with this edge.
    \end{enumerate}
\end{definition}

Any tree decomposition can be turned into a nice one of the same width in $\mathcal{O}(w^2 \cdot n)$ time, where $w$ is the width~\cite[Lemma 7.4, see also the discussion on page 168]{CyganFKLMPPS15} of the tree decomposition.

Moreover, we assume that the source $s=t$ is contained in all nodes of a nice tree decomposition $\T$ and that the root node, just as leaf nodes, contains only $s$ in the corresponding bag.

\subsection{Partial solutions}\label{partial_sol_and_naive_dp}

We will compute partial solutions for subgraphs corresponding to a node in a tree decomposition by combining partial solutions for subgraphs corresponding to its children. To make the definition more intelligible we introduce the following notation.

\begin{notation}[Subgraph induced by a node of a tree decomposition]
	Let $G=(V,E)$ be a graph and $\T=(T,\beta)$ be a nice tree decomposition for $G$ rooted in node $r$. For any node $x\in T$ we will denote by $G_x=(V_x,E_x)$ the subgraph induced by a tree decomposition node $x$ with
	\begin{itemize}
		\item $V_x = \bigcup_{y \text{ is a descendant of }x}B_{y}$,
		\item $E_x = \left\{e\in E\mid e\text{ is introduced in any descendant of }x\right\}$.
	\end{itemize}
	We consider that every $x\in T$ is a descendant of itself.
\end{notation}

With the necessary preparation, we can finally define a partial solution in subgraphs induced by a tree decomposition.

\begin{definition}
	Let $I=(G,\c,\w,s,t,\W)$ be a unified instance of WRP, $\T=(T,\beta)$ be a nice tree decomposition of $G$, $x\in T$. For every $X\subseteq B_x$ with $(\W \cap B_x)\subseteq X$ and every $L\subseteq X$ we call $S=(X,L)$ a \emph{presignature} at $x$. For a presignature $S=(X,L)$ at $x$ and partition $\Pttn$ of $X$ we call $(X,L,\Pttn)$ a \emph{solution signature} at~$x$.

	A subgraph $H\subseteq G_x$ is a \emph{partial solution} compatible with solution signature~$(X,L,\Pttn)$ at $x$, if all the following conditions hold
	\begin{enumerate}[(i)]
		\item every already introduced waypoint $w\in W\cap V_x$ is present in $V(H)$,
		\item $V(H)\cap B_x = X$,
		\item a vertex $v\in V(H)$ has odd degree in $H$ if and only if $v\in L$ and
		\item for every connected component $C$ of $H$ we have $V(C)\cap B_x\not=\emptyset$ and $V(C)\cap B_x$ is in~$\Pttn$.
	\end{enumerate}
\end{definition}
The signature of a solution allows us to recognize partial solutions that are equivalent from the global perspective.

A standard dynamic programming algorithm would be working in a bottom-up manner and computing for every signature a minimal weight partial solution. The computation runs from leaf nodes to the root node with the assumption that before computing an optimal partial solution of inner node $x$ all child nodes were already processed.

After completion of this procedure we ask for the weight computed for signature $(\{s\},\emptyset,\{\{s\}\})$ at the root $r$. It is easy to verify that a subgraph of $G=G_r$ is compatible with this signature if and only if it is connected, even, and contains all waypoints. By \Cref{lem:solution_subgraphs}, such a subgraph exists if and only if there is a solution walk of the same cost. 

This approach finds an optimal solution, but the running time of this procedure is suboptimal since we have to store the weight of an optimal partial solution for every signature, and the number of partitions is large.
As a first step, we store for each presignature only a subset of partitions for which there is a compatible partial solution together with the minimum weight of such a solution.
However, this is not sufficient to reduce the number of considered partitions significantly.
Hence, we describe the algorithm using the framework of the next subsection, which then allows us to easily speed the algorithm up.

\subsection{Manipulating Partitions}\label{sec:man_part}

We use the framework of Bodlaender et al.~\cite{BodlaenderCKN15} to improve the algorithm. This allows us, instead of considering all possible partitions, to limit ourselves to the \emph{representative sets} of a \emph{weighted partitions} containing all the needed information, as described later.
For each presignature $(X,L)$ we store a list of pairs $(\Pttn,w)$, where~$\Pttn$ is a partition of $X$ and $w$ is the weight of an optimal partial solution for solution signature $(X,L,\Pttn)$.

Let us first recall some notions concerning partitions.
Recall that \emph{a partition} of a set $U$ is a family of its subsets $\Pttn \subseteq 2^U \setminus \emptyset$ such that $\bigcup_{A \in \Pttn}A=U$ and for every two distinct $A,B \in \Pttn$ we have $A \cap B=\emptyset$. We denote by $\Pi(U)$ the set of all partitions of $U$. If $\Pttn$ is a partition, then we call $A \in \Pttn$ \emph{a block}. For $V \subseteq U$ we denote by $U[V]$ the partition of $U$ where all blocks are singletons except block $V$, i.e., $U[V]=\{V\}\cup \{\{v\}\mid v \in U \setminus V\}$. 

For $\Pttn \in \Pi(U)$ we denote by $\Pttn_{\downarrow V}$ the partition obtained by removing from $\Pttn$ all $x\not\in V$, i.e., $\Pttn_{\downarrow V}=\{ A \cap V \mid A \in \Pttn\} \setminus \emptyset$, and in a similar way we let $\Pttn_{\uparrow V}$ be the partition obtained from $\Pttn$ by adding singletons for every $x\in V\setminus U$, i.e, $\Pttn_{\uparrow V}=\Pttn \cup \{\{v\}\mid v \in V \setminus U\}$. For $\Pttn, \Qttn \in \Pi(U)$ we say that $\Pttn$ is a coarsening of $\Qttn$, denoted $\Qttn  \sqsubseteq \Pttn$, if each block of $\Qttn$ is contained in one block of $\Pttn$. The set $\Pi(U)$ together with coarsening relation $\sqsubseteq$ forms a lattice. We denote the join operation in this lattice $\sqcup$. In other words, $\Pttn  \sqcup \Qttn$ is the finest common coarsening of partitions $\Pttn$ and $\Qttn$. Note that the maximum element of the described lattice is $\{U\}$ and the minimum element is the partition of all $x\in U$ as singletons.

	A set of \emph{weighted partitions} is a subset $\A\subseteq\WPS$. A pair~$(\Pttn,w)$ is called weighted partition and $w$ its weight.

\begin{definition}[Operators on sets of weighted partitions~\cite{BodlaenderCKN15}]\label{def:operators}
	Let $U, U'$ be a sets, $\A\subseteq\WPS$, $\mathcal{B}\subseteq\Pi(U')\times\N$ sets of weighted partitions and $\rmc(\A)$ a set obtained by removing all duplicates from $\A$ while preserving only the one with minimal weight, i.e., \[\rmc(\A) := \left\{(\Pttn,w)\in\A \mid \nexists (\Pttn,w')\in\A \land w' < w\right\}.\] 
	We define the following operators.
	\begin{description}[leftmargin=2\parindent,labelindent=1\parindent]
		\item[\textbf{Union}] $\A\mincup\mathcal{B} := \rmc(\A\cup\mathcal{B})$ combine two weighted partitions sets $\A,\mathcal{B}$ and discard duplicates with larger weights. It can be used only if $U=U'$.
		\item[\textbf{Insert}] $\ins(V,\A) := \left\{(\Pttn_{\uparrow U\cup V},w)\mid(\Pttn, w)\in\A\right\}$, where $V\cap U=\emptyset$, expand the universe $U$ by $V$ and insert each $v\in V$ as a new partition $\{v\}$ into~$\Pttn$.
		\item[\textbf{Shift}] $\shift(w',\A) := \left\{(\Pttn, w + w')\mid(\Pttn, w)\in\A\right\}$ increase the weight of each partition by $w'\in\N$.
		\item[\textbf{Glue}] $\glue(\{u,v\},\A) := \rmc\left(\left\{(\hat{U}[\{u,v\}]\sqcup\Pttn_{\uparrow\hat{U}},w)\mid(\Pttn, w)\in\A\right\}\right)$, where ${\hat{U} =  U\cup\{u,v\}}$, for each partition merge sets containing $u$ and $v$ into one and add $u,v$ into $U$ if needed. By $\glue_\w(\{u,v\},\A)$ we denote $\shift(\omega(\{u,v\}),\glue(\{u,v\},\A))$.
		\item[\textbf{Project}] $\proj(V,\A) := \rmc\left(\left\{ (\Pttn_{\downarrow U\setminus V},w) \mid (\Pttn,w)\in\A \land \forall v\in V\,\exists v'\in U\setminus V\colon \Pttn \sqsupseteq U[\{v,v'\}] \right\}\right)$ remove every $v\in V$ from $U$ and from each partition while discarding each partition where the number of sets is decreased.
		\item[\textbf{Join}] $\join(\A,\mathcal{B}) := \rmc\left(\left\{ ( \Pttn_{\uparrow\hat{U}} \sqcup \mathcal{Q}_{\uparrow\hat{U}}, w_1 + w_2 )\mid(\Pttn,w_1)\in\A \land (\mathcal{Q},w_2)\in\mathcal{B} \right\}\right)$, where $\hat{U} = U\cup U'$, extend all partitions to the same set $\hat{U}$ and for each pair $(p,w_1)\in\A$ and $(q,w_2)\in\mathcal{B}$ return the result of the $\sqcup$ operation with weight equal to the sum of the weights of the original pairs.  
	\end{description}
\end{definition}

\subsection{The algorithm}

We will now formalize our dynamic programming approach using the operators from \Cref{sec:man_part} which allows us to later optimize the running time of the algorithm.
It is sufficient to describe how the algorithm processes in the different kinds of a nice tree decomposition nodes.

\paragraph{Leaf node $x$} The computation of the algorithm starts in leaf nodes. 
The subgraph corresponding to a leaf of the decomposition has single vertex $s$ and no edge at all, so the only valid presignature is $(\{s\},\emptyset)$ with the only possible partition being $\{\{s\}\}$. Thus
\[
	\A_x(\{s\},\emptyset) = \Big\{\big(\{\{s\}\},0\big)\Big\}.
\]

\paragraph{Introduce vertex $v$ node $x$ with child $y$} In an introduce vertex $v$ node $x$ of a tree decomposition $\T$ the presignature $(X,L)$ either prescribes to use the introduced vertex $v$ or not to use it. In the first case, we can be sure, thanks to the definition of a nice tree decomposition, that there is no edge that starts or ends in $v$ in~$G_x$.
Hence there is no partial solution in which $v$ has odd degree. Thus, if $v \in L$ then the set of weighted partitions is empty.
For any presignature $(X,L)$ such that $v\in X$ and $v \notin L$ the solution signature can be obtained from solution signature $(X\setminus\{v\}, L, \Pttn_y)$ of child node $y$ as $(X,L,\Pttn_y\cup\{\{v\}\})$. For the case $v \notin X$ the partial solutions do not change. Hence
\[
	\A_x(X,L) = \begin{cases}
			\ins(\{v\}, \A_y(X\setminus\{v\}, L)) & \text{if }v\in X\text{ and }v\not\in L, \\
			\A_y(X,L) & \text{if }v\not\in X\cup L, \\
			\emptyset & \text{otherwise.}
		\end{cases}
\]

\paragraph{Introduce edge $e=\{u,v\}$ node $x$ with child $y$} When node $x$ introduces edge $e=\{u,v\}$, the edge may or may not be added to partial solutions of child node~$y$. The easier part is when we decide to not add the edge to partial solutions, in that case, all the partial solutions are simply copied. On the other hand, if we add edge $e$ into the solution, then for every partial solution of child node $y$ the following happens
\begin{itemize}
 \item the weight $\w(e)$ of edge $e$ is added to the total weight of the partial solution,
 \item if $u$ does not belong to the same block as $v$, then both blocks are merged,
 \item if $u\not\in L$, then $u$ is added to $L$, the same for $v$,
 \item if $u\in L$, then $u$ is removed from $L$, the same for $v$.
\end{itemize}
In terms of presented operations, it yields
\[
	\A_x(X,L) = \begin{cases}
			\A_y(X,L) \mincup \glue_\w(\{u,v\}, \A_y(X,L\bigtriangleup\{u,v\})) & \text{if $u,v\in X$}\\
            \A_y(X,L) & \text{otherwise,}
		\end{cases}
\]
where $A\bigtriangleup B= (A\setminus B) \cup (B\setminus A)$ is the symmetric difference of the sets $A$ and~$B$.

\paragraph{Forget vertex $v$ node $x$ with child $y$} If vertex $v$ is forgotten in node $x$, then we must eliminate all the partial solutions of child node $y$ where $v$ is a singleton in~$\Pttn$ because there is no possibility to connect it with the remaining parts, and such partial solutions would not fulfill condition (iv) for partial solutions. By condition (iii) we can also eliminate partial solutions in which $v$ has odd degree. Thus
\[
\A_x(X,L) = \A_y(X,L)\mincup\proj(\{v\},\A_y(X\cup\{v\},L)).
\]

\paragraph{Join node $x$ with children $y$ and $z$} In join node $z$ we must properly combine the partial solutions of its children, which is exactly the purpose of the $\join$ operation. Hence,
\[
\A_x(X,L) = \bigmincup_{\substack{L_y,L_z\\ L = L_y\bigtriangleup L_z}} \mathtt{join}(\A_y(X,L_y), \A_z(X,L_z)).
\]

Let us now show that the recurrences correctly compute $\A_x(S)$ for every node $x$ and presignature $S$. This is captured by the following lemmas.

\begin{lemma}\label{lem:val_sol}
	For every node $x$, presignature $(X,L)$, and weighted partition $(\Pttn, w) \in \A_x(X,L)$ there is a partial solution $H$ compatible with $(X,L,\Pttn)$ at $x$ of cost at most $w$.
\end{lemma}
\begin{proof}
	We prove the lemma by a bottom-up induction on the decomposition tree. For a leaf node the only presignature is $(\{s\},\emptyset)$ with weighted partition $(\{\{s\}\},0)$ for which $(\{s\},\emptyset)$ is the partial solution. Let us now suppose that the lemma holds for every child node of node $x$. 
	
	If node $x$ with single child $y$ introduces vertex $v$, then for each presignature where $v\notin X$ and $v\notin L$ the partial solution compatible with $(X,L,\Pttn)$ at $y$ is also compatible with $(X,L,\Pttn)$ at $x$ and such a solution exists by the induction hypothesis. When $v\in X$ and $v\notin L$, then let $H'$ be a partial solution compatible with $(X\setminus\{v\},L,\Pttn\setminus\{\{v\}\})$ at $y$. Such a solution exists by the induction hypothesis. Then $H$, created by adding $v$ as an isolated vertex into  $H'$, is a partial solution compatible with $(X,L,\Pttn)$ at $x$. Any other presignature is invalid.
	
	When node $x$ with child $y$ introduces edge $\{u,v\}$ two cases can occur. The first case, when we decide to not add an edge $\{u,v\}$ to partial solutions valid for node $y$ is obvious. In the other case for solution signature $(X,L,\Pttn)$ where $u,v\in X$ we obtain a solution by adding edge $\{u,v\}$ to submultigraph $H'$ compatible with presignature $(X,L\bigtriangleup\{u,v\})$ in $y$. The weight of such a partial solution is increased by $\omega(\{u,v\})$.
	
	Let $x$ denote a forget vertex $v$ node with child $y$ and $(\Pttn, w) \in \A_x(X,L)$. If  $(\Pttn, w) \in \A_y(X,L)$, then there is a partial solution of weight at most $w$ compatible with $(X,L,\Pttn)$ at $y$. As it does not contain $v$, it is also compatible with $(X,L,\Pttn)$ at $x$. If $(\Pttn, w) \in \proj(v,\A_y(X\cup\{v\},L))$, then let $(\Pttn', w)$ be the weighted partition such that $(\Pttn, w) \in \proj(v,\{(\Pttn', w)\})$. We know that $\{v\} \notin \Pttn'$, as otherwise this partition would be discarded by $\proj$. By the induction hypothesis there is a partial solution $H$ of weight at most $w$ compatible with $(X\cup\{v\},L,\Pttn')$ at $y$. As the component containing $v$ also contains other vertices of $B_x$, $H$ is also compatible with $(X,L,\Pttn)$ at $x$.
	
	The last not yet discussed case is when node $x$ is a join node with exactly two children $y$, $z$. For $(\Pttn, w) \in \A_x(X,L)$ let $L_y,L_z$ be such that $(\Pttn, w) \in \mathtt{join}(\A_y(X,L_y), \A_z(X,L_z))$. Furthermore, let $(\Pttn_y, w_y) \in \A_y(X,L_y)$ and $(\Pttn_z, w_z) \in \A_z(X,L_z)$ be such that $(\Pttn, w) \in \mathtt{join}(\{(\Pttn_y, w_y)\}, \{(\Pttn_z, w_z)\})$. By the induction hypothesis there are partial solutions $H_y$ of weight at most $w_y$ compatible with $(X,L_y,\Pttn_y)$ at $y$ and $H_z$ of weight at most $w_z$ compatible with $(X,L_z,\Pttn_z)$ at $z$. Let $H= H_y \cup H_z$. We claim that $H$ is compatible with $(X,L,\Pttn)$ at $x$. Note that, as every edge is introduced exactly once, we have $E_y \cap E_z =\emptyset$, thus $E(H_y) \cap E(H_z) =\emptyset$ and the weight of $H$ is at most $w=w_y+w_z$. Clearly $V(H)\cap B_x =X$. As $V_x=V_y \cup V_z$, every waypoint $w \in \W \cap V_x$ is either in $\W \cap V_y$ or $\W \cap V_z$ and, hence, in $V(H) =V(H_y) \cup V(H_z) \supseteq (\W \cap V_y) \cup (\W \cap V_z)$.
	As $E(H_y) \cap E(H_z) =\emptyset$, a vertex has odd degree in $H$ if and only if it has odd degree in exactly one of $H_y$ and $H_z$, which is exactly if it is in $L=L_y \bigtriangleup L_z$. It follows from the properties of $\mathtt{join}$ that for each connected component $C$ of $H$ we have $C \cap B_x$ is in $\Pttn$. 
\end{proof}

\begin{lemma}\label{lem:sol_val}
	For every node $x$, presignature $(X,L)$, and weighted partition $\Pttn$ of $X$ such that there is a partial solution $H$ compatible with $(X,L, \Pttn)$ at $x$ of cost $w$, there is a pair $(\Pttn, w')$ in $\A_x(X,L)$ with $w' \le w$.
\end{lemma}
\begin{proof}
	We prove the lemma by a bottom-up induction on the tree decomposition similarly as in the proof of \Cref{lem:val_sol}.
	
	For a leaf node $x$ the only valid presignature is $(\{s\},\emptyset)$ and the only partition is $\{\{s\}\}$. Moreover $G_x = (\{s\},\emptyset)$ and, as $s \in W$, $H = (\{s\},\emptyset)$ is the only partial solution compatible with $(\{s\},\emptyset, \{\{s\}\})$ at $x$.
	Moreover, this partial solution has weight $0$, we stored $(\{\{s\}\},0)$ in $\A_x(\{s\},\emptyset)$, and, thus, for leaf nodes the lemma holds. 
	
	Let us now suppose that the lemma holds for every child node of node $x$ of the decomposition tree.
	
	When node $x$ with single child node $y$ introduces vertex $v$, then for any presignature $(X,L)$ and weighted partition $\Pttn$ of $X$ such that there is a partial solution $H$ compatible with $(X,L,\Pttn)$ we must distinguish two cases. 
	If the introduced vertex $v$ is not part of $H$, then $H \subseteq G_y$ and it is compatible with $(X,L,\Pttn)$ at $y$. 
	Hence, by the induction hypothesis we have $(\Pttn,w') \in \A_y(X,L)$ for some $w'\le w$. As $v \notin X$ and, thus, $\A_x(X,L)=\A_y(X,L)$ in this case, we have also $(\Pttn,w') \in \A_y(X,L)$. 
	If $v \in V(H)$, then, since the vertex $v$ is introduced in node $x$, there is no edge incident with $v$ in $G_x$ and also in $H$ which implies that $v\notin L$ and $v$ forms a connected component of $H$, and thus also a singleton partition in $\Pttn$. 
	Therefore $H'= H \setminus \{v\}$ is compatible with $(X\setminus\{v\},L,\Pttn')$ at $y$, where $\Pttn'$ is obtained from $\Pttn$ by removing the part $\{v\}$. Moreover, $H'$ has also weight $w$, as there are no edges incident on $v$ in $H$.
	Thus, by the induction hypothesis, $(\Pttn',w') \in \A_y(X\setminus\{v\},L)$ for some $w' \le w$. 
	Since $\A_x(X,L)=\ins(\{v\}, \A_y(X\setminus\{v\}, L))$ in this case, $(\Pttn,w')$ appears in $\A_x(X,L)$, as required.

	Let $x$ denote introduce edge $e=\{u,v\}$ node with single child node $y$ and $H$ a partial solution of weight $w$ compatible with solution signature $(X,L,\Pttn)$ at $x$. As in the previous type of tree decomposition node we must analyze several cases. 
	If $e\in E(H)$, then there is a partial solution $\widehat{H} := (V(H),E(H)\setminus\{e\})$ of weight $\widehat{w} = w - \w(e)$ compatible with $(X,L\bigtriangleup\{u,v\},\widehat{\Pttn})$ at $y$ for some $\widehat{\Pttn}$.
	Note that all connected components of $H$ are also connected components of $\widehat{H}$ except for the component containing $u$ and $v$.
	Therefore, the partition $\widehat{\Pttn}$ is such that all parts of $\widehat{\Pttn}$ containing neither $u$ nor $v$ also appear in $\Pttn$, if $u$ and $v$ are in the same part of $\widehat{\Pttn}$, then it also appears in $\Pttn$ and otherwise, the union of the parts of $\widehat{\Pttn}$ containing $u$ and $v$ appears in $\Pttn$.
	By the induction hypothesis, there is a pair $(\widehat{\Pttn},w')$ in $\A_y(X,L\bigtriangleup\{u,v\})$ such that $w'\leq \widehat{w}$. 
	Therefore, there is a pair $(\Pttn,w'')$ in $\glue_\w(e,\A_y(X,L\bigtriangleup\{u,v\}))$, such that $w''=w'+\w(e) \le \widehat{w}+\w(e)=w$.
	If $e\notin E(H)$, then $H \subseteq G_y$ and $H$ is compatible with $(X,L,\Pttn)$ at $y$. Therefore, by the induction hypothesis, the pair $(\Pttn,w'')$ is in $\A_y(X,L)$ for some $w'' \le w$.
	If $\{u,v\} \subseteq X$, then in both cases we get that $\A_x(X,L) =\A_y(X,L) \mincup \glue_\w(\{u,v\}, \A_y(X,L\bigtriangleup\{u,v\}))$ contains a pair $(\Pttn,\tilde{w})$ for some $\tilde{w} \le w''\le w$. Otherwise we have $\A_x(X,L) =\A_y(X,L)$ and the lemma follows.
		
	Let $x$ be a forget vertex $v$ node with a single child node $y$. On one hand, when $v$ is not a part of $H$, then $H$ is also a partial solution compatible with solution signature $(X,L,\Pttn)$ at~$y$. 
	Hence $\A_y(X,L)$ contains a pair $(\Pttn,w')$ for some $w' \le w$ by the induction hypothesis.
	On the other hand, if $v$ is a part of $H$, then $v$ must be in a connected component of $H$ containing some other vertex $u$ of $B_x$. 
	Therefore, $H$ is also compatible with signature $(X\cup\{v\},L,\Pttn')$ at $y$, where $\Pttn'$ is obtained from $\Pttn$ by adding $v$ to the part containing $u$. Thus, by the induction hypothesis, the pair $(\Pttn',w')$ is in $\A_y(X\cup\{v\},L)$ for some $w' \le w$. Hence, the pair $(\Pttn,w')$ is in $\proj(v,\A_y(X\cup\{v\},L))$. In either case, $\A_x(X,L) = \A_y(X,L)\mincup\proj(v,\A_y(X\cup\{v\},L))$ contains a pair $(\Pttn,\tilde{w})$ for some $\tilde{w} \le w' \le w$. 
		
	The last type of a node in a nice tree decomposition is a join node $x$ with exactly two children $y$ and $z$. For a given solution signature $(X,L,\Pttn)$ and a compatible partial solution $H$ we let $H_y=H \cap G_y$ and $H_z=H \cap G_z$ and let $w_y$ and $w_z$ be their weights, respectively.  
	We have that $H_y$ and $H_z$ are disjoint subgraphs, $H = H_y \cup H_z$, and, thus, $w_y + w_z = w$.
	We let $L_y$ ($L_z$) be the set of vertices of $H_y$ ($H_z$) of odd degree, respectively.
	For a vertex $v \in V(H) \cap (V_y \setminus V_z)$ all edges of $H$ incident on $v$ are in $H_y$, and, hence, $v \notin L_y$. Similarly for a vertex in $v \in V(H) \cap (V_z \setminus V_y)$. Thus $L_y \subseteq X$, $L_z \subseteq X$ and $L = L_y\bigtriangleup L_z$. 
	Also $H_y$ is compatible with $(X,L_y,\Pttn_y)$ at $y$ for some $\Pttn_y$ and $H_z$ is compatible with $(X,L_z,\Pttn_z)$ at $z$ for some $\Pttn_z$ such that $\Pttn_y \sqcup \Pttn_z =\Pttn$.
	Then, by the induction hypothesis, $\A_y(X,L_y)$ contains a pair $(\Pttn_y, w'_y)$ for some $w'_y \le w_y$ and $\A_z(X,L_z)$ contains a pair $(\Pttn_z, w'_z)$ for some $w'_z \le w_z$.
	Therefore $\mathtt{join}(\A_y(X,L_y), \A_z(X,L_z))$ contains a pair $(\Pttn, w')$ for some $w' \le w'_y+w'_z\le w_y+w_z=w$. Hence, also $\A_x(X,L) = \bigmincup_{\substack{L_y,L_z\\ L = L_y\bigtriangleup L_z}} \mathtt{join}(\A_y(X,L_y), \A_z(X,L_z))$ contains a pair $(\Pttn, w'')$ for some $w'' \le w' \le w$ and the lemma holds.
\end{proof}

\subsection{Using Representative Sets}

In order to speed up the algorithm, we do not compute the whole set $\A_x$ for each node $x$ and each presignature, but just its suitable subset. We need the following definitions and theorems.

\begin{definition}[Representation and representative set]
	For a set of weighted partitions $\A\subseteq\WPS$ and partition $q\in\Pi(U)$, let
	\[
		\opt(q,\A):=\min\left(\left\{w\mid(p,w)\in\A\land p\sqcup q = \{U\}\right\} \cup \{\infty\}\right).
	\]
	
	For any other set $\A'\subseteq\WPS$ of weighted partitions we say that $\A'$ is a \emph{representative set} for $\A$ if and only if
	\[
	\forall q\in\Pi(U)\colon \opt(q,\A') = \opt(q,\A).
	\]
	
	In addition, we say that a function $f\colon2^{\WPS}\times Z\to2^{\WPS}$, where $Z$ denotes any combination of further inputs, \emph{preserves representation} if for every $\A,\A'\in\WPS$ and every $z\in Z$ it holds that if $\A'$ is a representative set for~$\A$ then $f(\A',z)$ remains a representative for $f(\A,z)$.
\end{definition}

\begin{theorem}[Properties of Operators~\cite{BodlaenderCKN15}]\label{thm:operators}
The operators described in \Cref{def:operators} preserve representation.
Furthermore, all of them excluding $\mathtt{join}$ can be executed in $s\cdot|U|^{\mathcal{O}(1)}$ time, where $s$ is the size of operation input. The running time of the $\join$ operation is $|\A|\cdot|\mathcal{B}|\cdot|U|^{\mathcal{O}(1)}$.
\end{theorem}

\begin{theorem}[Bodlaender et al.~\cite{BodlaenderCKN15}]\label{reduce}
	There is an algorithm for the $\reduce$ operation that given a set of weighted partitions $\A\subseteq\WPS$ produces in time $2^{(\omega-1)|U|}\cdot|U|^{\mathcal{O}(1)}\cdot|\A|$, where $\omega < 2.373$ is a matrix multiplication exponent, a representative subset $\A'\subseteq\A$ such that $|\A'|\leq2^{|U|-1}$.
\end{theorem}

The above theorem states that a suitably small representative set of weighted partitions can be always found in a reasonable time. This completes the introduction of the used toolbox.

In order to speed up the algorithm, we do not compute the whole set $\A_x$ for each node $x$ and each presignature, but just its representative subset $\A'_x$ as follows.
For each node $x$ and its presignature $(X,L)$ we first compute a set $\A''_x(X,L)$ based on $\A'_y$ (and possibly $\A'_z$), using the recurrences described in the previous section.
Then we use algorithm $\reduce$ from \Cref{reduce} to compute representative set $\A'_x(X,L)$ of $\A''_x(X,L)$, such that $|\A'_x(X,L)|\leq 2^{\mathcal{O}(\tw)}$.
Since in the recurrences we use only operators from \Cref{def:operators} which all preserves representation by \Cref{thm:operators}, set $\A_x'(X,L)$ remains representative for $\A_x(X,L)$.

\begin{theorem}
	There exist an algorithm that given an instance $(G,\c,\w, s,t,\W)$ of the \WRP{} solves it in $2^{\mathcal{O}(\tw)}\cdot n$ time.
\end{theorem}
\begin{proof}
    Our algorithm first preprocesses the instance to obtain a unified one. This increases the number of vertices by at most one and the number of edges is at most 2 more than twice the original number. Also the treewidth is increased by at most 1. 
    Since the number of edges is $\mathcal{O}(\tw \cdot n)$~\cite[Exercise 7.15]{CyganFKLMPPS15}, the unification can be performed in $\mathcal{O}(\tw \cdot n)$ time.
    For simplicity, we denote the unified instance also by $(G,\c,\w, s,t,\W)$. 
    Then we obtain a tree decomposition for $G$ using the algorithm of Bodlaender et al.~\cite{BodlaenderDDFLP16}, turn it into a nice one along the lines of \cite[Lemma 7.4, see also p. 168]{CyganFKLMPPS15}, and finally add $s$ to every bag. We obtain a nice tree decomposition of width $\mathcal{O}(\tw)$ with $\mathcal{O}(\tw^{\mathcal{O}(1)}n)$ nodes and $s$ in every bag in time $2^{\mathcal{O}(\tw)}\cdot n$.
    
    Thereafter the introduced dynamic programming algorithm computes the representative set $\A'_x(X,L)$ of $\A_x(X,L)$ for every node $x$ of the tree decomposition and every presignature $(X,L)$ at $x$. By \Cref{lem:val_sol} and \Cref{lem:sol_val}, a pair $(\{\{s\}\},w)$ is in $\A_r(\{s\}, \emptyset)$ if and only if there is a solution of weight $w$. Since in the recurrence we use only operators from \Cref{def:operators} which all preserves representation, set $\A_x'$ remains representative for all nodes $x$ including the root node~$r$. Hence $\A'_r(\{s\}, \emptyset)$ is representative for $\A_r(\{s\}, \emptyset)$, which necessarily implies $\A_r(\{s\}, \emptyset)=\A'_r(\{s\}, \emptyset)$ and the algorithm is correct.
    
    Let us discuss the running time of the proposed algorithm. As the width of the used decomposition is $\mathcal{O}(\tw)$, there are $2^{\mathcal{O}(\tw)}$ presignatures for each bag. We use the $\reduce$ procedure to obtain each $\A'_x(X,L)$, hence we have $|\A'_x(X,L)|\leq 2^{|X|-1} \le 2^{\mathcal{O}(\tw)}$. In each of the recurrences, we use a constant number of operators from \Cref{def:operators}. The most time consuming is the computation in join nodes which takes at most $|\A_y(X,L_y)| \cdot |\A_z(X,L_z)| \cdot |X|^{\mathcal{O}(1)} = 2^{|X|-1} \cdot 2^{|X|-1} \cdot |X|^{\mathcal{O}(1)} = 2^{\mathcal{O}(\tw)}$ time. Therefore the bottleneck of our algorithm is the $\reduce$ procedure. Nevertheless, since the size of intermediate sets of weighted partitions never exceeds $2^{\mathcal{O}(\tw)}$, this procedure also runs in $2^{\mathcal{O}(\tw)}$ time. As the nice tree decomposition has $\mathcal{O}(\tw^{\mathcal{O}(1)}n)$ nodes, we can say that the overall running time of the presented algorithm is $2^{\mathcal{O}(\tw)}\cdot n$. 
\end{proof}

\section{ETH Lower Bound}\label{sec:lb}

The Exponential Time Hypothesis (ETH for short) introduced by Impagliazzo and Paturi~\cite{ImpagliazzoP01} states that there is a constant $\delta_3>0$ such that there is no algorithm for 3-SAT with running time $2^{\delta_3 n}m^{\mathcal{O}(1)}$, where $n$ is the number of variables and $m$ the total length of the input formula. To prove the optimality of our algorithm, we will use the following ETH implication.

\begin{theorem}[{Impagliazzo, Paturi, and Zane~\cite{ImpagliazzoPZ01}, see also~\cite[Theorem 14.6]{CyganFKLMPPS15}}]\label{thm:HamCyc}
	Unless ETH fails, \textsc{Hamiltonian Cycle} admits no algorithm working in $2^{o(n+m)}$ time, where $n$ and $m$ are the number of vertices and edges of the input graph, respectively. 
\end{theorem}

Using the preceding theorem, we are able to easily prove the following.

\begin{theorem}
	Unless ETH fails, there is no algorithm for the \WRP working in $2^{o(n+m)}$ time and in particular none working in $2^{o(\tw(G))}\cdot n^{\mathcal{O}(1)}$  time, where $n$, $m$, and $\tw(G)$ are the number of vertices, edges, and the treewidth of the input graph, respectively. 
\end{theorem}
\begin{proof}
	Assume to the contrary that such an algorithm $\mathbb{A}$ exists. Note that, since $\tw(G) \le n-1$ for any graph $G$, we have $2^{o(\tw(G))}\cdot n^{\mathcal{O}(1)} \subseteq 2^{o(n)} \subseteq 2^{o(n+m)}$. We show that the existence of $\mathbb{A}$ contradicts \autoref{thm:HamCyc}. 

	Consider the following algorithm. Given an undirected graph $G=(V,E)$---an instance of \textsc{Hamiltonian Cycle}---let $\c(e)=\w(e)=1$ for every $e \in E$, $\W=V$, and $s=t$ be an arbitrary vertex in $V$. Now run $\mathbb{A}$ on $(G,\c,\w,s,t,\W)$. If the returned walk has weight $|V|$, then accept $G$, otherwise reject it. As a Hamiltonian cycle in $G$ corresponds to a walk of length $|V|$ visiting each vertex of $\W$ exactly once, using each edge at most once, and starting in $s=t$ and vice versa, the answer is correct. 

	Hence the running time of the algorithm contradicts \autoref{thm:HamCyc}, finishing the proof.
\end{proof}

\section{Conclusion}

We presented a deterministic algorithm for the \WRP in undirected graphs running in $2^{\mathcal{O}(\tw)}\cdot n$ time. An interesting open problem is to determine the complexity of the problem in directed graphs with underlying undirected graphs of small treewidth. While the correspondence with degree constrained submultigraphs via Eulerian trails is still valid in directed graphs, it is no longer true that each edge is traversed at most twice in an optimal walk. It is easy to find instances in which a particular edge must be traversed as many as $n-1$ times, where $n$ is the number of vertices, in any feasible walk. Furthermore, while in the final submultigraph the indegree of each vertex must be equal its outdegree, for a partial solution the difference between these two degrees can be arbitrarily large. This makes the problem more challenging in directed graphs.

\bibliography{references}

\end{document}